\documentclass[11pt,openany]{article}
\usepackage{latexsym}
\usepackage{amssymb}
\usepackage{epsfig}
\usepackage{graphicx}
\usepackage{bm}
\usepackage[mathscr]{eucal}
\renewcommand{\theequation}{\thesection.\arabic{equation}}
\begin{document}

\title{{\Large {\bf Unitary limit and quantum interference effect in disordered two-dimensional
crystals with nearly half-filled bands}}}
\author{Y.~H.~Yang,$^{1,2}$ D.~Y.~Xing,$^2$ Y.~G.~Wang,$^1$ and  M.~Liu,$^1$\\
\em\small $^{1}$Department of Physics, Southeast University,
Nanjing 210096, China\\
\em\small $^{2}$National Laboratory of Solid State
Microstructures, Nanjing University, Nanjing 210008, China}
\date{}
\maketitle
\begin{abstract}
{\large Based on the self-consistent $T$-matrix approximation, the
quantum interference (QI) effect is studied with the diagrammatic
technique in weakly-disordered two-dimensional crystals with
nearly half-filled bands. In addition to the usual 0-mode cooperon
and diffuson, there exist $\pi$-mode cooperon and diffuson in the
unitary limit due to the particle-hole symmetry. The diffusive
$\pi$-modes are gapped by the deviation from the exactly-nested
Fermi surface. The conductivity diagrams with the gapped
$\pi$-mode cooperon or diffuson are found to give rise to
unconventional features of the QI effect. Besides the inelastic
scattering, the thermal fluctuation is shown to be also an
important dephasing mechanism in the QI processes related with the
diffusive $\pi$-modes. In the proximity of the nesting case, a
power-law anti-localization effect appears due to the $\pi$-mode
diffuson. For large deviation from the nested Fermi surface, this
anti-localization effect is suppressed, and the conductivity
remains to have the usual logarithmic weak-localization correction
contributed by the 0-mode cooperon. As a result, the dc
conductivity in the unitary limit becomes a non-monotonic function
of the temperature or the sample size, which is quite different
from the prediction of the usual weak-localization theory.}
\end{abstract}
\vspace{0.2in} PACS number(s): 73.20.Fz, 71.30.+h, 72.15.Rn,
73.20.Jc
\newpage

\begin{center}
{\section*{{\bf I. Introduction}}}
\end{center}
\setcounter{section}{1} \setcounter{equation}{0} \vspace{0.5cm}

Since the pioneering work of Anderson~\cite{1}, the
disorder-induced localization in electronic systems has been the
subject of intense research activities for more than four decades.
According to the scaling theory proposed by Abrahams et
al.~\cite{2}, all electronic states in one- and two-dimensional
(1D and 2D) disordered systems are localized irrespective of the
degree of the randomness, and in 3D systems there exist the
metal-insulator transitions. The prediction of the scaling theory
for 1D systems is in agreement with the exact results~\cite{3,4},
and the existence of Anderson localization in 3D systems is widely
accepted~\cite{5}. The dimension 2 is the marginal dimension in
the problem of Anderson localization. It is believed that in
generic situations, there exist no true metallic states in 2D
disordered noninteracting systems at zero temperature. However,
some exceptions have been known to this rule. For example,
non-localized states were found at the band center of a 2D
Anderson model with purely off-diagonal disorder~[6-10]. The
metal-insulator transition is also recently shown to occur in a
disordered 2D square lattice model with long-range power-law
transfer terms~\cite{11}. Therefore, the disorder effects in 2D
noninteracting electron systems deserve further investigation.

In the scaling theory of localization, the perturbative approach
is exploited to study the weak-disorder limit, revealing the
weak-localization effect in 2D electron systems~\cite{5,12,13}.
The weak-localization theory describes the quantum interference
(QI) effect that results from constructive interference between
the closed electron paths and their time-reversal counterparts,
which can be represented by the maximally-crossed diagrams in the
diagrammatic approach. While the QI effect contributes no singular
correction to the electronic density of states (DOS), the QI
correction to the dc conductivity of a 2D system is shown to
be~\cite{5,12,13}
\begin{equation}
\delta\sigma=-\frac{e^2}{2\pi^2}\times\left\{
\begin{array}{cc}
\ln(\gamma_0/\gamma_i),~&{\rm if}~\gamma_i\gg\gamma_L,\\
\ln(\gamma_0/\gamma_L),~&{\rm if}~\gamma_L\gg\gamma_i,
\end{array}\right.
\end{equation}
where $\gamma_0$ and $\gamma_i$ denote, respectively, the elastic
and inelastic scattering rates, and $\gamma_L=D/2L^2$ with $D$ the
diffusion coefficient and $L$ the sample size. At finite
temperatures, the inelastic scattering has a dephasing effect on
the QI process. Equation (1.1) indicates that the dc conductivity
is a monotonic function of the temperature or the sample size, as
the inelastic scattering rate $\gamma_i$ decreases with decreasing
the temperature.

Equation (1.1) is usually obtained by means of the free-electron
model within the Born approximation, in which only the twofold
scattering process from the same impurity is concerned. However,
for a 2D crystal with half-filled band, there usually exist the
particle-hole symmetry of electronic spectrum, as well as the van
Hove singularity in the DOS at the Fermi level. The particle-hole
symmetry has been recently shown to play an important role on the
QI effect of the quasiparticles in 2D disordered $d$-wave
superconductors~\cite{14,15}, as well as on the disorder effects
in 2D Mott-Hubbard insulators~\cite{16}. The large value of the
DOS at the Fermi level can drive a disordered system into the
unitary limit~\cite{17}. Therefore, it is highly desirable to
investigate the effects of the particle-hole symmetry and van Hove
singularity on the QI process in disordered 2D crystals. While
both the Born and unitary limits have been widely investigated in
the disordered $d$-wave superconductors, the QI effect in the
unitary limit did not receive enough attention in the study of
disordered normal metals.

In the previous short report~\cite{18}, we have studied the QI
effect at zero temperature in disordered 2D crystals with exactly
half-filled bands. We pointed out that it is necessary to treat
the impurity scattering within the self-consistent $T$-matrix
approximation (SCTMA) due to the existence of the van Hove
singularity at the Fermi level. The SCTMA approach takes into
account the effect of multiple-scattering events from the same
impurity, and thus gives naturally the results of the Born and the
unitary approximations in two distinct limits. In addition to the
usual 0-mode cooperon and diffuson, there exist the $\pi$-mode
cooperon and diffuson in the unitary limit due to the
particle-hole symmetry. These diffusive $\pi$-modes were found to
introduce some new conductivity diagrams, and the resulting QI
correction to the conductivity in the unitary limit was shown to
increase with the square of the sample size, suggesting the
existence of the extended states at the band center.

The aim of this work is to study further the situation of
deviation from exact half-filling, and investigate the dephasing
effect on the QI process at finite temperatures. Much details of
the derivations omitted in Ref.~[18] is also presented in this
paper. Contrary to the 0-mode cooperon and diffuson, the diffusive
$\pi$-modes are found to be gapped by the deviation from the
exactly-nested Fermi surface. The conductivity diagrams with these
gapped diffusive $\pi$-modes give rise to unexpected features of
the QI effect. In addition to the inelastic scattering, the
thermal fluctuation is shown to be also an important dephasing
mechanism in the QI processes related with the diffusive
$\pi$-modes. Upon approaching to the nested Fermi surface, the
conductivity is subject to a power-law anti-localization
correction induced by the $\pi$-mode diffuson. For large deviation
from the nesting case, the contributions of the diffusive
$\pi$-modes are suppressed, and the usual logarithmic
weak-localization effect remains due to the 0-mode cooperon.
Consequently, the QI correction to the conductivity in the unitary
limit has a non-monotonic behavior with increasing the temperature
or the sample size. This result is strikingly different from the
prediction of the usual weak-localization theory.

The structure of this paper is as follows. In Sec.~II, the SCTMA
scenario is described in details for a disordered 2D crystal with
nearly half-filled band, and the conditions for the Born and
unitary limits are presented. Based on the SCTMA, the expressions
of the 0-mode and $\pi$-mode cooperons and diffusons are derived
in Sec.~III for the nearly-nested Fermi surface. In Sec.~IV, we
evaluate the electric current-current correlation function
responsible for the QI effect, and then calculate the QI
correction to dc conductivity in the unitary limit. The
conclusions are summarized in Sec.~V. In Appendix A, we prove some
mathematical formulas, which are useful in the evaluations of the
previous sections. Appendix B is presented to show that the
electric current-current correlation function in the
retarded-retarded (RR) channel has a vanishing QI correction to
the dc conductivity.

\vspace{0.5cm}
\begin{center}
{\section*{{\bf II. Self-consistent $T$-matrix approximation}}}
\end{center}
\setcounter{section}{2} \setcounter{equation}{0} \vspace{-0.8cm}

We start with the tight-binding model for a square lattice, of
which the electrical spectrum is described by
\begin{equation}
\epsilon_{\bm k}=-2t(\cos k_xa+\cos k_ya)-\mu,
\end{equation}
where $a$ is the lattice constant, $t$ is the nearest-neighbor
hopping integral, and $\mu$ is the chemical potential. A
half-filled band ($\mu=0$) is characteristic of a perfectly-nested
Fermi surface, which is composed of four straight lines ($k_x\pm
k_y=\pi/a$ and $k_x\pm k_y=-\pi/a$) in the momentum space. The
electronic velocity ${\bm v}_{\bm k}=\nabla\epsilon_{\bm k}$ is
given by
\begin{equation}
v^\alpha_{\bm k}=2ta\sin k_\alpha a,~(\alpha=x,y),
\end{equation}
here we take $\hbar=k_{\rm B}=1$. The DOS per spin of the clean
crystal, defined by $\rho_{\rm cl}(\epsilon)=\sum_{\bm
k}\delta(\epsilon-\epsilon_{\bm k})$, is shown to be of a
logarithmic van Hove singularity at the center of the band, i.e.,
\begin{equation}
\rho_{\rm
cl}(\epsilon)=\frac{1}{2\pi^2a^2t}\ln\frac{16t}{|\epsilon+\mu|},~\Big({\rm
for}~|\epsilon+\mu|\ll 4t\Big).
\end{equation}

In the presence of dilute point-like nonmagnetic impurities
randomly substituted for the host atoms, the electronic
self-energy can be adequately expressed in the SCTMA as
\begin{equation}
\Sigma^{R(A)}=n_iT^{R(A)}=\eta\gamma_0\mp i\gamma_0,
\end{equation}
where $n_i$ is the impurity concentration, $\gamma_0$ is the
elastic scattering rate, $\eta$ is a dimensional parameter, and
$\eta\gamma_0$ represents the shift of the chemical potential that
can be absorbed in $\mu$. A use of Dyson's equation immediately
yields the impurity-averaged one-particle Green's functions as
\begin{equation}
G^{R(A)}_{\bm k}(\epsilon)=\frac{1}{\epsilon-\epsilon_{\bm k}\pm
i\gamma},
\end{equation}
where $\gamma=\gamma_0+\gamma_i$, with $\gamma_i$ the additional
inelastic scattering rate induced by, e.g., the electron-electron
and/or electron-phonon interactions. The intent of the
introduction of $\gamma_i$ is to investigate the dephasing effect
on the QI process at finite temperatures that are assumed to be
low enough so that $\gamma_i(T)\ll \gamma_0$. The DOS at the Fermi
level of this disordered  crystal can be readily calculated via
\begin{equation}
\rho_0=-\frac{1}{\pi}{\rm Im}\sum_{\bm k}G^R_{\bm
k}=\frac{1}{\pi}\sum_{\bm k}\frac{\gamma}{\epsilon_{\bm
k}^2+\gamma^2},
\end{equation}
with $G^{R(A)}_{\bm k}=G^{R(A)}_{\bm k}(0)$. We consider,
throughout this paper, the nearly half-filled band and the case of
weak disorder, meaning that $|\mu|\ll \gamma\ll t$. Then,
substituting Eq.~(A1) in Appendix A into Eq.~(2.6), we obtain
\begin{equation}
\rho_0=\frac{1}{2\pi^2a^2t}\ln\frac{16t}{\gamma}.
\end{equation}

The elastic scattering rate $\gamma_0$ and the parameter $\eta$
can be evaluated consistently by means of the $T$-matrix equation
\begin{equation}
T^{R(A)}=V+V\sum_{\bm k}G^{R(A)}_{\bm k}T^{R(A)},
\end{equation}
with $V$ the strength of the impurity potential. A combination of
Eq.~(2.4) with Eq.~(2.8) leads to
\begin{equation}
\gamma_0=\frac{n_i}{\pi\rho_0(1+\eta^2)}
\end{equation}
and
\begin{equation}
\eta=\frac{1}{\pi\rho_0V}.
\end{equation}
The Born limit corresponds to $\eta^2\gg 1$, leading to
\begin{equation}
\gamma_0=\pi n_i\rho_0V^2;
\end{equation}
and the unitary limit corresponds to $\eta^2\ll 1$, yielding
\begin{equation}
\gamma_0=\frac{n_i}{\pi\rho_0}.
\end{equation}

Equations (2.11) and (2.12) have been pointed out in
Ref.~\cite{17}. The Born and unitary approximations turns out to
be two natural limits of the SCTMA, as the latter includes the
effect of multiple-scattering process from the same impurity. As
shown by Eq.~(2.7), $\rho_0$ has a large value in the
weak-disorder case ($\gamma$ is very small). From Eq.~(2.10), it
follows that the unitary condition can be satisfied for a finite
impurity potential $V$ and a low impurity concentration $n_i$.

\begin{center}
{\section*{{\bf III. The diffusive modes}}}
\end{center}
\setcounter{section}{3} \setcounter{equation}{0} \vspace{0.5cm}

Since the QI effect in a disordered electron system results from
the cooperon and diffuson, we first derive their expressions for a
nearly half-filled band on the basis of the SCTMA described above.
It is useful to note that the present system has two different
kinds of symmetries. The first is the time-reversal symmetry
described by
\begin{equation}
\epsilon_{-\bm k}=\epsilon_{\bm k},~{\bm v}_{-\bm k}=-{\bm v}_{\bm
k},
\end{equation}
and
\begin{equation}
G^{R(A)}_{-\bm k}(\epsilon)=G^{R(A)}_{\bm k}(\epsilon),
\end{equation}
due to which the cooperon and diffuson have the same expressions.
The other symmetry is given by
\begin{equation}
\epsilon_{{\bm Q}+{\bm k}}=-\epsilon_{\bm k}-2\mu,~{\bm v}_{{\bm
Q}+{\bm k}}=-{\bm v}_{\bm k},
\end{equation}
and
\begin{equation}
G^{R(A)}_{{\bm Q}+{\bm k}}(\epsilon)=-G^{A(R)}_{\bm
k}(-\epsilon-2\mu),
\end{equation}
where ${\bm Q}$ is one of four nesting vectors
$(\pm\pi/a,\pm\pi/a)$. This symmetry is termed as particle-hole
symmetry for the case of nested Fermi surface, and gives rise to
the existence of the $\pi$-mode cooperon and diffuson in the
unitary limit~\cite{18}. The ladder diagrams for the $\pi$-mode
cooperon are depicted in Fig.~1.
\begin{figure}[htbp]
  \begin{center}
  \psfig{file=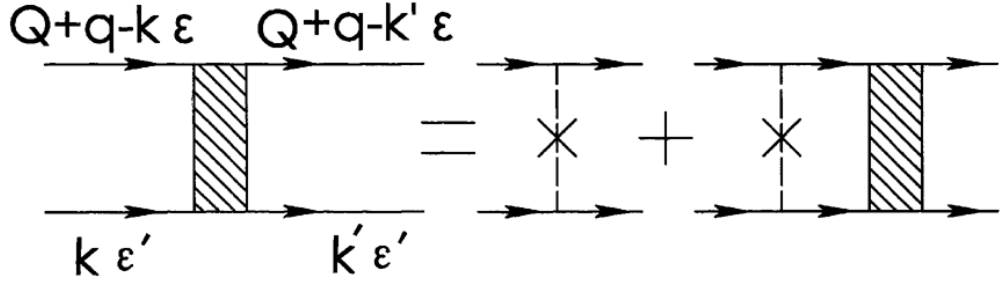,width=14cm,bbllx=186pt,bblly=400pt,bburx=405pt,bbury=477pt}
    \parbox{12.5cm}{\caption{\footnotesize Ladder diagrams for the $\pi$-mode cooperon (shaded
blocks). The dashed lines with crosses represent the $T$-matrix
elements of the impurity scattering. ${\bm Q}$ denotes the nesting
vector. The $\pi$-mode diffuson corresponds to the similar
diagrams with the arrows of the upper particle lines inverted. The
diagrams for the 0-mode cooperon and diffuson can be obtained by
setting ${\bm Q}=0$ in those of the diffusive $\pi$-modes.  }}
       \end{center}
\end{figure}
Unlike in the Born approximation used in the usual
weak-localization theory, the dashed lines in Fig.~1 denote the
$T$-matrix elements instead of the impurity potential $V$.

The equation for 0-mode cooperon in the retarded-advanced (RA)
channel is given by
\begin{equation}
C({\bm q};\epsilon,\epsilon')^{RA}=W^{RA}+W^{RA}H({\bm
q};\epsilon,\epsilon')^{RA}C({\bm q};\epsilon,\epsilon')^{RA},
\end{equation}
where
\begin{equation}
W^{RA}=n_iT^RT^A=\frac{\gamma_0}{\pi\rho_0}
\end{equation}
and
\begin{equation}
H({\bm q};\epsilon,\epsilon')^{RA}=\sum_{\bm k}G^R_{{\bm q}-{\bm
k}}(\epsilon)G^A_{\bm k}(\epsilon').
\end{equation}
Expanding the right side of Eq.~(3.7) in powers of small ${\bm
q}$, $\epsilon$, and $\epsilon'$, we obtain
\begin{equation}
H({\bm q};\epsilon,\epsilon')^{RA}=\sum_{\bm k}\Big[G^R_{\bm
k}G^A_{\bm k}+\frac{1}{2}{\bm q}{\bm q}:(\nabla\nabla G^R_{\bm
k})G^A_{\bm k}-\epsilon(G^R_{\bm k})^2G^A_{\bm
k}-\epsilon'G^R_{\bm k}(G^A_{\bm k})^2\Big].
\end{equation}
By carrying out a partial integral for the second term in the
right side of Eq.~(3.8), and using Eqs.~(A2) and (A3), one gets
\begin{eqnarray}
H({\bm q};\epsilon,\epsilon')^{RA}&=&\sum_{\bm k}\Big[G^R_{\bm
k}G^A_{\bm k}-\frac{1}{4}q^2v^2_{\bm k}(G^R_{\bm k}G^A_{\bm
k})^2-\epsilon(G^R_{\bm k})^2G^A_{\bm k}-\epsilon'G^R_{\bm
k}(G^A_{\bm
k})^2\Big]\nonumber\\&=&\frac{\pi\rho_0}{2\gamma^2}\Big[2\gamma-
Dq^2+i(\epsilon-\epsilon')\Big],
\end{eqnarray}
where the diffusion coefficient $D$ is given by
\begin{equation}
D=\frac{2t}{\pi^2\gamma\rho_0}.
\end{equation}
Substituting Eqs.~(3.6) and (3.9) into Eq.~(3.5), and noting that
the cooperon and diffuson have the same expressions, we obtain
\begin{equation}
C({\bm q};\epsilon,\epsilon')^{RA}=D({\bm
q};\epsilon,\epsilon')^{RA}=\frac{2\gamma^2}{\pi\rho_0}\frac{1}
{Dq^2-i(\epsilon-\epsilon')+2\gamma_i}.
\end{equation}
The 0-mode cooperon and diffuson in the RR channel are
non-singular, given by
\begin{equation}
C({\bm q};\epsilon,\epsilon')^{RR}=D({\bm
q};\epsilon,\epsilon')^{RR}=-\frac{\gamma_0}{\pi\rho_0}.
\end{equation}
Equations (3.11) and (3.12) are suitable for all values of
$\eta^2$ for the present crystal, while they were previously
obtained in the Born limit for the free-electron
model~\cite{5,12,13}.

As in the disordered $d$-wave superconductors~\cite{14,15}, the
$\pi$-mode cooperon and diffuson exist only in the unitary limit
($\eta^2\ll1$). The equation for the $\pi$-mode cooperon in the RR
channel is given by
\begin{equation}
C_\pi({\bm q};\epsilon,\epsilon')^{RR}=W^{RR}+W^{RR} H_\pi({\bm
q};\epsilon,\epsilon')^{RR}C_\pi({\bm q};\epsilon,\epsilon')^{RR},
\end{equation}
where
\begin{equation}
W^{RR}=n_i(T^R)^2=-\frac{\gamma_0}{\pi\rho_0},~({\rm
for}~\eta^2\ll 1)
\end{equation}
and
\begin{equation}
H_\pi({\bm q};\epsilon,\epsilon')^{RR}=\sum_{\bm k}G^R_{{\bm
Q}+{\bm q}-{\bm k}}(\epsilon)G^R_{\bm k}(\epsilon').
\end{equation}
Substituting Eq.~(3.4) into Eq.~(3.15), one can easily show that
\begin{equation}
H_\pi({\bm
q};\epsilon,\epsilon')^{RR}=-\frac{\pi\rho_0}{2\gamma^2}\Big[2\gamma-
Dq^2+i(\epsilon+\epsilon'+2\mu)\Big].
\end{equation}
A substitution of Eqs.~(3.14) and (3.16) into Eq.~(3.13)
immediately yields
\begin{equation}
C_\pi({\bm q};\epsilon,\epsilon')^{RR(AA)}=D_\pi({\bm
q};\epsilon,\epsilon')^{RR(AA)}=-\frac{2\gamma^2}{\pi\rho_0}\frac{1}{Dq^2
\mp i(\epsilon+\epsilon'+2\mu)+2\gamma_i}.
\end{equation}
Similarly, we have
\begin{equation}
C_\pi({\bm q};\epsilon,\epsilon')^{RA}=D_\pi({\bm
q};\epsilon,\epsilon')^{RA}=\frac{\gamma_0}{\pi\rho_0}.
\end{equation}
We stress that the existence of the diffusive $\pi$-modes in RR or
AA channel stems from the unitary condition and the particle-hole
symmetry, similarly with the situation of disordered $d$-wave
superconductors~\cite{14,15}.

It is instructive to compare the diffusive poles of the 0-mode and
$\pi$-mode cooperons (diffusons) for zero temperature
($\gamma_i=0$). The 0-mode cooperon in RA channel is gapless,
having a diffusive pole at a small energy difference and a small
total momentum. The $\pi$-mode cooperon in RR or AA channel is
gapped by the deviation from the perfectly-nested Fermi surface,
with the diffusive pole at a total energy of $-2\mu$ and a total
momentum of the nesting vector. As will be shown below, the above
difference between the 0-mode and $\pi$-mode cooperons (diffusons)
makes the QI effect in the unitary limit significantly different
from that of the Born limit.

\vspace{0.5cm}
\begin{center}
{\section*{{\bf IV. QI correction to the conductivity}}}
\end{center}
\setcounter{section}{4} \setcounter{equation}{0}\vspace{-0.5cm}

The above expressions for the cooperon and difffuson can be used
to calculate the QI correction to the conductivity. The mean free
path of the electrons is calculated as
\begin{equation}
l=\sqrt{\frac{D}{2\gamma}}=\frac{\sqrt{2}ta}{\gamma\ln(16t/\gamma)},
\end{equation}
which is much longer than the lattice constant $a$ for the
considered situation ($\gamma\ll t$). Therefore, the usual
quasiclassical approach can be used to calculate the conductivity.
The electrical current density is expressed by ${\bm
j}=-e\sum_{{\bm k}\sigma}{\bm v}_{\bm k}c^\dagger_{{\bm
k}\sigma}c_{{\bm k}\sigma}$, with $c^\dagger_{{\bm k}\sigma}$ and
$c_{{\bm k}\sigma}$ standing for the creation and annihilation
operators of electrons, respectively. According to the Kubo
formula, the dc conductivity related with the impurity scattering
is given by~\cite{19}
\begin{equation}
\sigma=\sigma^{RA}+\sigma^{RR},
\end{equation}
with
\begin{eqnarray}
\sigma^{RA(RR)}&=&\pm\lim_{\omega\rightarrow0}\frac{1}{\omega}
\int\limits_{-\infty}^{+\infty}\frac{d\epsilon}{2\pi}
[f(\epsilon-\omega)-f(\epsilon)]{\rm
Re}\Pi(\epsilon,\epsilon-\omega)^{RA(RR)}\nonumber\\
&=&\pm\int\limits_{-\infty}^{+\infty}\frac{d\epsilon}{2\pi}
\frac{{\rm
Re}\Pi(\epsilon,\epsilon)^{RA(RR)}}{4T\cosh^2(\epsilon/2T)},
\end{eqnarray}
where $f(\epsilon)$ denotes the Fermi distribution function, and
$\Pi(\epsilon,\epsilon')^{RA}$ and $\Pi(\epsilon,\epsilon')^{RR}$
represent the electrical current-current correlation functions in
RA and RR channels, respectively. In this paper, only the case of
\begin{equation}
\gamma_i\ll\gamma~{\rm and}~T\ll\gamma,
\end{equation}
is considered.

The Drud conductivity corresponds to the contribution of the
``bare bubble" diagram, which is given by (for $T=0$)
\begin{equation}
\sigma_0=\frac{1}{2\pi}{\rm
Re}\Pi(0,0)^{RA}_0=\frac{e^2}{2\pi}\sum_{\bm k}v^2_{\bm k}G^R_{\bm
k}G^A_{\bm k}.
\end{equation}
Using Eq.~(A3), we obtain
\begin{equation}
\sigma_0=2e^2\rho_0D,
\end{equation}
meaning that the Einstein relation is also suitable for the
present nearly half-filled band.

\begin{center} {\subsection*{\bf A. Correlation functions responsible for the QI effect}}
\end{center}

The correlation functions responsible for the QI effect can be
calculated by the polarization diagrams with cooperon and
diffuson. As in the usual weak-localization theory, the 0-mode
diffuson has no contribution to the QI effect. The contribution of
0-mode cooperon to the conductivity is represented by Fig.~2(c)
\begin{figure}[htbp]
  \begin{center}
  \psfig{file=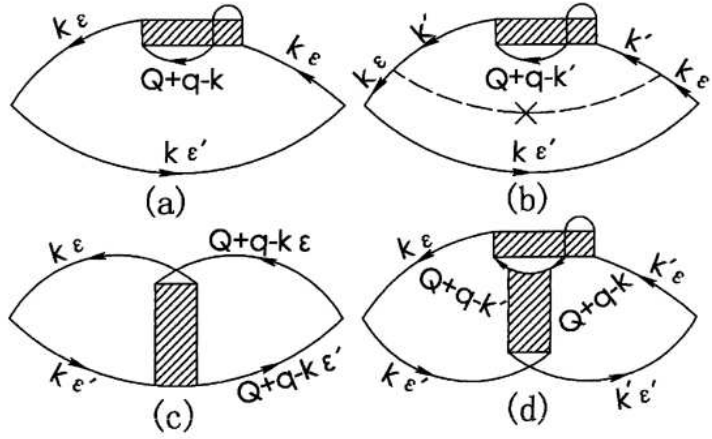,width=10cm,bbllx=226pt,bblly=393pt,bburx=365pt,bbury=485pt}
    \parbox{12.5cm}{\caption{\footnotesize Leading conductivity diagrams with $\pi$-mode cooperon
(shaded blocks).  }}
       \end{center}
\end{figure}
with ${\bm Q}=0$ (the maximally-crossed diagrams), and the
corresponding correlation function can be evaluated as
\begin{eqnarray}
\Pi(\epsilon,\epsilon')^{RA}_{\rm coop}&=&e^2\sum_{{\bm k}{\bm
q}}{\bm v}_{\bm k}\cdot{\bm v}_{-\bm k}G^R_{\bm k}G^R_{-\bm
k}G^A_{\bm k}G^A_{-\bm k}C({\bm
q};\epsilon,\epsilon')^{RA}\nonumber\\&=&-e^2\sum_{{\bm k}{\bm
q}}v^2_{\bm k}(G^R_{\bm k}G^A_{\bm k})^2C({\bm
q};\epsilon,\epsilon')^{RA}.
\end{eqnarray}
Substituting Eq.~(3.11) into Eq.~(4.7), and using Eq.~(A3), we get
\begin{equation}
\Pi(\epsilon,\epsilon')^{RA}_{\rm coop}=-\sum_{\bm
q}\frac{4e^2D}{Dq^2-i(\epsilon-\epsilon')+2\gamma_i},
\end{equation}
which is valid for all values of $\eta^2$.

Besides the 0-mode cooperon, the $\pi$-mode cooperon and diffuson
also have contributions to the conductivity in the unitary limit,
and the leading polarization diagrams are shown in Figs.~2 and 3,
respectively~\cite{18}. These diagrams can be generated using the
conserving approximation, as in the theory of disordered
interacting electron systems~\cite{5,20,21}. In Appendix B, we
show that the RR-correlation functions from the diagrams in
Figs.~2 and 3 contribute vanishing corrections to the dc
conductivity, similarly with the case of disordered interacting
electrons~\cite{22}. This feature is a manifestation of the
conserving law for the number of particles. Thus, only the
RA-correlation functions are calculated in this section.

The non-trivial contributions of the $\pi$-mode cooperon to the
RA-correlation function result from Figs.~2(a) and 2(b). They are
expressed, respectively, as
\begin{eqnarray}
\Pi(\epsilon,\epsilon')^{RA}_{2a}&=&e^2\sum_{{\bm k}{\bm q}}{\bm
v}_{\bm k}\cdot{\bm v}_{\bm k}(G^R_{\bm k})^2G^R_{{\bm Q}-{\bm
k}}G^A_{\bm k}C_\pi({\bm
q};\epsilon,\epsilon)^{RR}\nonumber\\&=&-e^2\sum_{{\bm k}{\bm
q}}v^2_{\bm k}(G^R_{\bm k}G^A_{\bm k})^2C_\pi({\bm
q};\epsilon,\epsilon)^{RR}
\end{eqnarray}
and
\begin{eqnarray}
\Pi(\epsilon,\epsilon')^{RA}_{2b}&=&e^2\sum_{{\bm k}{\bm k'}{\bm
q}}{\bm v}_{\bm k}\cdot{\bm v}_{\bm k}(G^R_{\bm k}G^R_{\bm
k'})^2G^R_{{\bm Q}-{\bm k'}}G^A_{\bm k}W^{RR}C_\pi({\bm
q};\epsilon,\epsilon)^{RR}\nonumber\\&=&-e^2\sum_{{\bm k}{\bm
k'}{\bm q}}v^2_{\bm k}(G^R_{\bm k}G^R_{\bm k'})^2G^A_{\bm
k}G^A_{\bm k'}W^{RR}C_\pi({\bm q};\epsilon,\epsilon)^{RR},
\end{eqnarray}
where the particle-hole symmetry shown by Eq.~(3.4) has been used.
Substituting Eqs.~(3.14) and (3.17) into Eqs.~(4.9) and (4.10),
and using Eqs.~(A2) and (A3), one can readily show that
\begin{equation}
\Pi(\epsilon,\epsilon')^{RA}_{2a}=2\Pi(\epsilon,\epsilon')^{RA}_{2b}=\sum_{\bm
q}\frac{4e^2D}{Dq^2-i2(\epsilon+\mu)+2\gamma_i}.
\end{equation}

\begin{figure}[htbp]
  \begin{center}
  \psfig{file=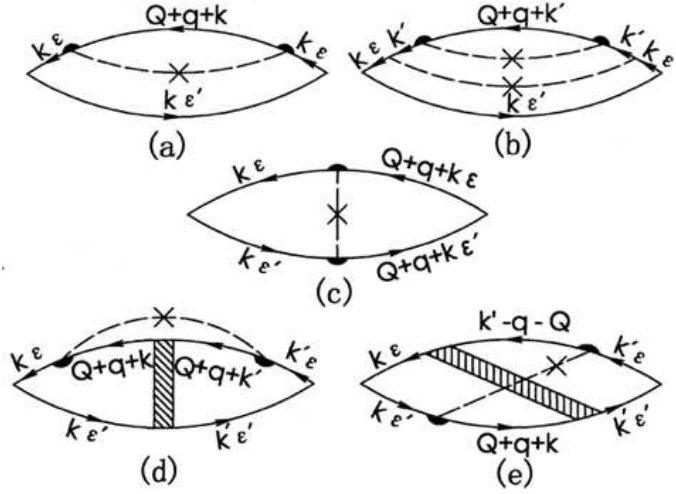,width=10cm,bbllx=230pt,bblly=390pt,bburx=358pt,bbury=485pt}
    \parbox{12.5cm}{\caption{\footnotesize Leading conductivity diagrams with $\pi$-mode diffuson
(shaded blocks). The dark parts denote the vertex corrections to
the impurity scattering due to the $\pi$-mode diffuson.  }}
       \end{center}
\end{figure}
The contributions of Figs.~3(a)--3(c) to the RA-correlation
function can be expressed, respectively, as
\begin{eqnarray}
\Pi(\epsilon,\epsilon')^{RA}_{3a}&=&e^2\sum_{{\bm k}{\bm q}}{\bm
v}_{\bm k}\cdot{\bm v}_{\bm k}(G^R_{\bm k})^2G^R_{{\bm Q}+{\bm
k}}G^A_{\bm k}W^{RR}\Lambda^R({\bm
q},\epsilon)^2\nonumber\\&=&-e^2\sum_{{\bm k}{\bm q}}v^2_{\bm k}
(G^R_{\bm k}G^A_{\bm k})^2W^{RR}\Lambda^R({\bm q},\epsilon)^2,
\end{eqnarray}
\begin{eqnarray}
\Pi(\epsilon,\epsilon')^{RA}_{3b}&=&e^2\sum_{{\bm k}{\bm k'}{\bm
q}}{\bm v}_{\bm k}\cdot{\bm v}_{\bm k}(G^R_{\bm k}G^R_{\bm k'})^2
G^R_{{\bm Q}+{\bm k'}}G^A_{\bm k}(W^{RR})^2\Lambda^R({\bm
q},\epsilon)^2\nonumber\\&=&-e^2\sum_{{\bm k}{\bm k'}{\bm q}}
v^2_{\bm k}(G^R_{\bm k}G^R_{\bm k'})^2G^A_{\bm k}G^A_{\bm
k'}(W^{RR})^2\Lambda^R({\bm q},\epsilon)^2,
\end{eqnarray}
and
\begin{eqnarray}
\Pi(\epsilon,\epsilon')^{RA}_{3c}&=&e^2\sum_{{\bm k}{\bm q}}{\bm
v}_{\bm k}\cdot{\bm v}_{{\bm Q}+{\bm k}}G^R_{\bm k}G^R_{{\bm
Q}+{\bm k}}G^A_{{\bm Q}+{\bm k}}G^A_{\bm k}W^{RA}\Lambda^R({\bm
q},\epsilon)\Lambda^A({\bm
q},\epsilon')\nonumber\\&=&-e^2\sum_{{\bm k}{\bm q}}v^2_{\bm
k}(G^R_{\bm k}G^A_{\bm k})^2W^{RA}\Lambda^R({\bm
q},\epsilon)\Lambda^A({\bm q},\epsilon'),
\end{eqnarray}
where $\Lambda^{R(A)}({\bm q},\epsilon)$ represent the vertex
functions of the impurity scattering due to the $\pi$-mode
diffuson, as shown by Fig.~4.
\begin{figure}[htbp]
  \begin{center}
  \psfig{file=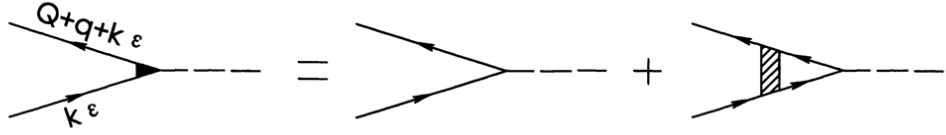,width=13cm,bbllx=176pt,bblly=415pt,bburx=414pt,bbury=455pt}
    \parbox{12.5cm}{\caption{\footnotesize Diagrams for the vertex correction to the impurity
scattering. The shaded block stands for the $\pi$-mode diffuson.
}}
       \end{center}
\end{figure}
Obviously, the retarded vertex function is given by
\begin{equation}
\Lambda^R({\bm q},\epsilon)=1+D_\pi({\bm
q};\epsilon,\epsilon)^{RR}H_\pi({\bm q};\epsilon,\epsilon)^{RR}.
\end{equation}
A substitution of Eqs.~(3.16) and (3.17) into Eq.~(4.15)
immediately yields
\begin{equation}
\Lambda^{R(A)}({\bm q},\epsilon)=\frac{2\gamma}{Dq^2\mp
i2(\epsilon+\mu)+2\gamma_i},
\end{equation}
for $Dq^2\ll\gamma$ and $|\epsilon|\ll\gamma$. Substituting
Eqs.~(3.6), (3.14), and (4.16) into Eqs.~(4.12)--(4.14), and using
Eqs.~(A2) and (A3), we obtain
\begin{equation}
\Pi(\epsilon,\epsilon')^{RA}_{3a}=2\Pi(\epsilon,\epsilon')^{RA}_{3b}=\sum_{\bm
q}\frac{8e^2D\gamma}{[Dq^2-i2(\epsilon+\mu)+2\gamma_i]^2}
\end{equation}
and
\begin{equation}
\Pi(\epsilon,\epsilon')^{RA}_{3c}=-\sum_{\bm
q}\frac{8e^2D\gamma}{[Dq^2-i2(\epsilon+\mu)+2\gamma_i][Dq^2+i2(\epsilon'+\mu)+2\gamma_i]}.
\end{equation}

Similarly, the RA-correlation functions corresponding to
Figs.~3(d) and 3(e) are expressed by
\begin{equation}
\Pi(\epsilon,\epsilon')^{RA}_{3d}=e^2\sum_{{\bm k}{\bm k'}{\bm
q}}{\bm v}_{\bm k}\cdot{\bm v}_{\bm k'}G^R_{\bm k}G^R_{\bm
k'}G^R_{{\bm Q}+{\bm q}+{\bm k}}G^R_{{\bm Q}+{\bm q}+{\bm
k'}}G^A_{\bm k}G^A_{\bm k'}W^{RR}\Lambda^R({\bm
q},\epsilon)^2D_\pi({\bm q};\epsilon,\epsilon')^{RA}
\end{equation}
and
\begin{equation}
\Pi(\epsilon,\epsilon')^{RA}_{3e}=e^2\sum_{{\bm k}{\bm k'}{\bm
q}}{\bm v}_{\bm k}\cdot{\bm v}_{\bm k'}G^R_{\bm k}G^R_{\bm
k'}G^R_{-{\bm Q}-{\bm q}+{\bm k'}}G^A_{{\bm Q}+{\bm q}+{\bm
k}}G^A_{\bm k}G^A_{\bm k'}W^{RA}\Lambda^R({\bm
q},\epsilon)\Lambda^A({\bm q},\epsilon')D_\pi({\bm
q};\epsilon,\epsilon')^{RA}.
\end{equation}
For small ${\bm q}~(Dq^2\ll\gamma)$, Eqs.~(4.19) and (4.20) can be
changed as
\begin{equation}
\Pi(\epsilon,\epsilon')^{RA}_{3d}=\frac{e^2}{4}\sum_{{\bm k}{\bm
k'}{\bm q}}q^2v^2_{\bm k}v^2_{\bm k'}G^R_{\bm k}G^R_{\bm
k'}(G^A_{\bm k}G^A_{\bm k'})^3W^{RR}\Lambda^R({\bm
q},\epsilon)^2D_\pi({\bm q};\epsilon,\epsilon')^{RA}
\end{equation}
and
\begin{equation}
\Pi(\epsilon,\epsilon')^{RA}_{3e}=-\frac{e^2}{4}\sum_{{\bm k}{\bm
k'}{\bm q}}q^2v^2_{\bm k}v^2_{\bm k'}G^A_{\bm k}G^R_{\bm
k'}(G^R_{\bm k}G^A_{\bm k'})^3W^{RA}\Lambda^R({\bm
q},\epsilon)\Lambda^A({\bm q},\epsilon')D_\pi({\bm
q};\epsilon,\epsilon')^{RA}.
\end{equation}
Substituting Eqs.~(3.6), (3.14), (3.18), and (4.16) into
Eqs.~(4.21) and (4.22), and using Eqs.~(A2) and (A3), we obtain
\begin{equation}
\Pi(\epsilon,\epsilon')^{RA}_{3d}=-\sum_{\bm
q}\frac{e^2D^2q^2}{[Dq^2-i2(\epsilon+\mu)+2\gamma_i]^2},
\end{equation}
and
\begin{equation}
\Pi(\epsilon,\epsilon')^{RA}_{3e}=-\sum_{\bm
q}\frac{e^2D^2q^2}{[Dq^2-i2(\epsilon+\mu)+2\gamma_i][Dq^2+i2(\epsilon'+\mu)+2\gamma_i]}.
\end{equation}

The expressions for the correlation function, Eqs.~(4.8), (4.11),
(4.17), (4.18), (4.23), and (4.24), are the main results we obtain
in this subsection, and will be used to calculate the dependence
of the conductivity on the temperature or sample size. It is
worthy to point out that Eqs.~(4.17) and (4.18) are more divergent
than the others. The contributions of Figs.~3(a)--3(c) to the
conductivity have been presented in Ref.~\cite{18} for the case of
$\mu=T=0$. In the following calculations of the conductivity, we
will neglect the terms of the order $e^2/\pi^2$, which is much
smaller than the Drud one, i.e., $e^2/\pi^2\sigma_0=\gamma/4t\ll
1$.

\vspace{0.5cm}
\begin{center} {\subsection*{\bf B. Temperature dependence of the conductivity}}
\end{center}\vspace{-0.8cm}

In non-unitary cases, only the 0-mode cooperon contributes to the
QI effect. Substituting Eq.~(4.8) into Eq.~(4.3), and noting that
the upper cutoff of $q$ is set to be $1/l$, we can readily obtain
\begin{equation}
\delta\sigma(T)=\sigma(T)^{RA}_{\rm
coop}=-\frac{e^2}{2\pi^2}\ln\frac{\gamma}{\gamma_i},
\end{equation}
where we have used
\[
\int_0^{+\infty}\frac{dx}{\cosh^2x}=1.
\]
Although Eq.~(4.25) was obtained within the Born approximation in
the usual weak-localization theory, we show that it is valid for
all values of $\eta^2$ in the present system. The logarithmic
weak-localization effect results from the singular backscattering
due to the time-reversal symmetry $v_{-{\bm k}}=-v_{\bm k}$.

In the unitary limit, the $\pi$-mode cooperon and diffuson also
contribute to the conductivity. Substituting Eq.~(4.11) into
Eq.~(4.3), and completing the integral over ${\bm q}$, we get
\begin{equation}
\sigma(T)^{RA}_{\pi-{\rm
coop}}=2\sigma(T)^{RA}_{2a}+2\sigma(T)^{RA}_{2b}
=\frac{3e^2}{8\pi^2}\int\limits_{-\infty}^{+\infty}\frac{dx}{\cosh^2x}
\ln\frac{\gamma^2}{(2Tx+\mu)^2+\gamma_i^2},
\end{equation}
yielding
\begin{equation}
\sigma(T)^{RA}_{\pi-{\rm coop}}=\frac{3e^2}{2\pi^2}\times\left\{
\begin{array}{cc}
\ln(\gamma/\gamma_i),~&{\rm if}~\gamma_i\gg T,|\mu|,\\
\ln(\gamma/T),~&{\rm if}~T\gg\gamma_i,|\mu|,\\
\ln(\gamma/|\mu|),~&{\rm if}~|\mu|\gg T,\gamma_i.
\end{array}\right.
\end{equation}
Equation (4.27) indicates that the $\pi$-mode cooperon contributes
a logarithmic anti-localization correction to the conductivity.
This weak anti-localization correction occurs as the result of the
singular forward-scattering processes shown by Figs.~2(a) and
2(b).

Similarly, by means of Eqs.~(4.17), (4.18), (4.23), and (4.24), we
can show that
\begin{eqnarray}
\sigma(T)^{RA}_{3a}&=&2\sigma(T)^{RA}_{3b}=\frac{e^2}{4\pi^2}\int\limits_{-\infty}^{+\infty}
\frac{dx}{\cosh^2x}\frac{\gamma\gamma_i}{(2Tx+\mu)^2+\gamma_i^2}\nonumber\\
&=&\frac{e^2}{2\pi^2}\times\left\{
\begin{array}{cc}
\gamma/\gamma_i,~&{\rm if}~\gamma_i\gg T,|\mu|,\\
\pi\gamma/4T,~&{\rm if}~T\gg\gamma_i,|\mu|,\\
\gamma\gamma_i/\mu^2,~&{\rm if}~|\mu|\gg T,\gamma_i,
\end{array}\right.
\end{eqnarray}
\begin{eqnarray}
\sigma(T)^{RA}_{3c}&=&-\frac{e^2}{4\pi^2}\int\limits_{-\infty}^{+\infty}
\frac{dx}{\cosh^2x}\frac{\gamma}{2Tx+\mu}\arctan\frac{2Tx+\mu}{\gamma_i}\nonumber\\
&=&-\frac{e^2}{2\pi^2}\times\left\{
\begin{array}{cc}
\gamma/\gamma_i,~&{\rm if}~\gamma_i\gg T,|\mu|,\\
\pi\gamma/4T,~&{\rm if}~T\gg\gamma_i,|\mu|,\\
\pi\gamma/2|\mu|,~&{\rm if}~|\mu|\gg T,\gamma_i,
\end{array}\right.
\end{eqnarray}
and
\begin{eqnarray}
\sigma(T)^{RA}_{3d}&=&\sigma(T)^{RA}_{3e}
=-\frac{e^2}{32\pi^2}\int\limits_{-\infty}^{+\infty}\frac{dx}{\cosh^2x}
\ln\frac{\gamma^2}{(2Tx+\mu)^2+\gamma_i^2}\nonumber\\
&=&-\frac{e^2}{8\pi^2}\times\left\{
\begin{array}{cc}
\ln(\gamma/\gamma_i),~&{\rm if}~\gamma_i\gg T,|\mu|,\\
\ln(\gamma/T),~&{\rm if}~T\gg\gamma_i,|\mu|,\\
\ln(\gamma/|\mu|),~&{\rm if}~|\mu|\gg T,\gamma_i,
\end{array}\right.
\end{eqnarray}
where we have used
\[
\lim\limits_{x_0\to0^+}\int\limits_{-\infty}^{+\infty}
\frac{dx}{\cosh^2x}\frac{x_0}{x^2+x_0^2}=\pi,
\]
and
\[
\lim\limits_{x_0\to0^+}\int\limits_{-\infty}^{+\infty}
\frac{dx}{\cosh^2x}\frac{1}{x}\arctan\frac{x}{x_0}=\pi.
\]
Therefore, the contribution from the $\pi$-mode diffuson is given
by
\begin{eqnarray}
\sigma(T)^{RA}_{\pi-{\rm
diff}}&=&2\sigma(T)^{RA}_{3a}+2\sigma(T)^{RA}_{3b}+
\sigma(T)^{RA}_{3c}+2\sigma(T)^{RA}_{3d}+2\sigma(T)^{RA}_{3e}\nonumber\\
&=&\frac{e^2}{2\pi^2}\times\left\{
\begin{array}{cc}
2\gamma/\gamma_i-\ln(\gamma/\gamma_i),~&{\rm if}~\gamma_i\gg T,|\mu|,\\
\pi\gamma/2T-\ln(\gamma/T),~&{\rm if}~T\gg\gamma_i,|\mu|,\\
-\pi\gamma/2|\mu|-\ln(\gamma/|\mu|),~&{\rm if}~|\mu|\gg
T,\gamma_i.
\end{array}\right.
\end{eqnarray}
The dominant power-law terms in Eq.~(4.31) come from the more
divergent contributions of Figs.~3(a)--3(c). While Figs.~3(a) and
3(b) correspond to singular forward scatterings, Fig.~3(c)
describes a singular backscattering process due to the
particle-hole symmetry $v_{{\bm Q}+{\bm k}}=-v_{\bm k}$.

The total QI correction to the dc conductivity in the unitary
limit is given by
\[
\delta\sigma_{\tiny{\rm U}}(T)=\sigma(T)^{RA}_{\rm
coop}+\sigma(T)^{RA}_{\pi-{\rm coop}}+\sigma(T)^{RA}_{\pi-{\rm
diff}},
\]
having the following temperature behavior:
\begin{equation}
\delta\sigma_{\tiny{\rm U}}(T)\sim\frac{e^2}{2\pi^2}\times\left\{
\begin{array}{cc}
2\gamma/\gamma_i,~&{\rm if}~\gamma_i\gg T,|\mu|,\\
\pi\gamma/2T,~&{\rm if}~T\gg\gamma_i,|\mu|,\\
-\ln(\gamma/\gamma_i),~&{\rm if}~|\mu|\gg T,\gamma_i.
\end{array}\right.
\end{equation}
At low temperatures, it is reasonable to assume that
$\gamma_i=\gamma(T/T_0)^p$ with $p>1$ for the electron-electron or
electron-phonon interactions~\cite{5,12}. Then we get
\begin{equation}
\delta\sigma_{\tiny{\rm U}}(T)\sim\frac{e^2}{2\pi^2}\times\left\{
\begin{array}{cc}
2(T_0/T)^p,~&{\rm if}~\gamma_i\gg T,|\mu|,\\
\pi\gamma/2T,~&{\rm if}~T\gg\gamma_i,|\mu|,\\
-p\ln(T_0/T),~&{\rm if}~|\mu|\gg T,\gamma_i.
\end{array}\right.
\end{equation}
Equation (4.33) indicates that the QI correction to the dc
conductivity in the unitary limit is a non-monotonic function of
the temperature. The temperature dependence of the conductivity
changes from power-law anti-localization behaviors to a
logarithmic localization one with decreasing the temperature,
where the crossover occurs at $T\sim|\mu|$.

The temperature behavior of the conductivity can be understood by
analyzing the dephasing effects on the QI processes. While the QI
process related with the 0-mode cooperon is dephased by the
inelastic scattering, both the inelastic scattering and the
thermal fluctuation are the dephasing factors of the QI effects
resulting from the diffusive $\pi$-modes. This is because the
diffusive 0-modes and $\pi$-modes have different diffusive poles.
In the region of ${\rm min}\{\gamma_i,T\}\gg|\mu|$, the QI effect
results dominantly from the contributions of diagrams 3(a)--3(c)
that contain the $\pi$-mode diffuson. In the case of $\gamma_i\gg
T,|\mu|$, the inelastic scattering is the main dephasing
mechanism. For the situation of $T\gg\gamma_i,|\mu|$, however, the
thermal fluctuation becomes the dominant dephsing factor. If the
deviation from the nesting Fermi surface is large enough so that
$|\mu|\gg\gamma_i,T$, the contributions of the diffusive
$\pi$-modes are suppressed, and the QI effect stems only from the
0-mode cooperon that is dephased by the inelastic scattering.

\begin{center} {\subsection*{\bf C. Sample-size dependence of the conductivity}}
\end{center}

At zero temperature ($T=\gamma_i=0$), the conductivity formula,
Eq.(4.3), reduces to be
\begin{equation}
\sigma(L)^{RA(RR)}=\pm\frac{1}{2\pi}{\rm Re}\Pi(0,0)^{RA(RR)}.
\end{equation}
For a finite system, the upper and lower cutoffs of $q$ in the
correlation functions are set to be $1/l$ and $1/L$, respectively.

In non-unitary cases, the conductivity is easily shown to be
subject to a logarithmic weak-localization correction as
\begin{equation}
\delta\sigma(L)=\sigma(L)^{RA}_{\rm
coop}=-\frac{e^2}{2\pi^2}\ln\frac{\gamma}{\gamma_L},
\end{equation}
which is in agreement with Eq.~(1.1).

In the unitary limit, the $\pi$-mode cooperon is shown to yield a
logarithmic anti-localization correction to the conductivity as
\begin{eqnarray}
\sigma(L)^{RA}_{\pi-{\rm
coop}}&=&2\sigma(L)^{RA}_{2a}+2\sigma(L)^{RA}_{2b}=\frac{3e^2}{4\pi^2}\ln\frac{\gamma^2}
{\gamma_L^2+\mu^2}\nonumber\\
&=&\frac{3e^2}{2\pi^2}\times\left\{
\begin{array}{cc}
\ln(\gamma/\gamma_L),~&{\rm if}~\gamma_L\gg|\mu|,\\
\ln(\gamma/|\mu|),~&{\rm if}~|\mu|\gg\gamma_L.
\end{array}\right.
\end{eqnarray}
Similarly, substituting  Eqs.~(4.17), (4.18), (4.23), and (4.24)
into Eq.~(4.34), we can readily show that
\begin{equation}
\sigma(L)^{RA}_{3a}=2\sigma(L)^{RA}_{3b}
=\frac{e^2}{2\pi^2}\frac{\gamma\gamma_L}{\gamma_L^2+\mu^2},
\end{equation}
\begin{equation}
\sigma(L)^{RA}_{3c}=-\frac{e^2}{2\pi^2}\frac{\gamma}{\mu}\arctan\frac{\mu}{\gamma_L},
\end{equation}
and
\begin{equation}
\sigma(L)^{RA}_{3d}=\sigma(L)^{RA}_{3e}=-\frac{e^2}{16\pi^2}
\ln\frac{\gamma^2}{\gamma_L^2+\mu^2}.
\end{equation}
Thus, we get
\begin{eqnarray}
\sigma(L)^{RA}_{\pi-{\rm
diff}}&=&2\sigma(L)^{RA}_{3a}+2\sigma(L)^{RA}_{3b}+\sigma(L)^{RA}_{3c}
+2\sigma(L)^{RA}_{3d}+2\sigma(L)^{RA}_{3e}\nonumber\\
&=&\frac{e^2}{2\pi^2}\times\left\{
\begin{array}{cc}
2\gamma/\gamma_L-\ln(\gamma/\gamma_L),~&{\rm if}~\gamma_L\gg|\mu|,\\
-\pi\gamma/2|\mu|-\ln(\gamma/|\mu|),~&{\rm if}~|\mu|\gg\gamma_L.
\end{array}\right.
\end{eqnarray}
The dominant power-law terms in above equation are also from the
contributions of Figs.~3(a)--3(c).

As a result, the total QI correction to the conductivity in the
unitary limit is given by
\begin{eqnarray}
\delta\sigma_{\tiny\rm U}(L)&=&\sigma(L)^{RA}_{\rm
coop}+\sigma(L)^{RA}_{\pi-{\rm coop}} +\sigma(L)^{RA}_{\pi-{\rm
diff}}\nonumber\\&\sim &\frac{e^2}{2\pi^2}\times\left\{
\begin{array}{cc}
2\gamma/\gamma_L,~&{\rm if}~\gamma_L\gg|\mu|,\\
-\ln(\gamma/\gamma_L),~&{\rm if}~|\mu|\gg\gamma_L.
\end{array}\right.
\end{eqnarray}
It is instructive to define a characteristic length as
\begin{equation}
\xi=\sqrt{\frac{D}{2|\mu|}}=l\sqrt{\frac{\gamma}{|\mu|}}\gg l.
\end{equation}
Then Eq.~(4.41) can be expressed as
\begin{equation}
\delta\sigma_{\tiny\rm U}(L)\sim\frac{e^2}{\pi^2}\times\left\{
\begin{array}{cc}
L^2/l^2,~&{\rm if}~L\ll\xi,\\
-\ln(L/l),~&{\rm if}~L\gg\xi,
\end{array}\right.
\end{equation}
which is also a non-monotonic function of the sample size. For a
small sample ($L\ll\xi$), the conductivity increases with $L^2$
due to the contributions of diagrams 3(a)--3(c) containing the
$\pi$-mode diffuson. In the case of $L\gg\xi$, the QI corrections
from the diffusive $\pi$-modes are suppressed, and the usual
weak-localization effect remains due to the contribution of 0-mode
coooperon.

\vspace{0.5cm}
\begin{center}
{\section*{{\bf V. Conclusions}}}
\end{center}
\setcounter{section}{5} \setcounter{equation}{0} \vspace{-0.5cm}

Using the SCTMA, we have investigated theoretically the effect of
multiple-scattering of impurities on the QI of electrons in 2D
crystals with nearly half-filled bands. The particle-hole
symmetry, together with the van Hove singularity at the band
center, is shown to alter qualitatively the results predicted by
the usual weak-localization theory.

For a definite impurity potential $V$, the large value of $\rho_0$
( the DOS at the Fermi level) can drive the system into the
unitary limit with decreasing the impurity concentration $n_i$.
While the usual 0-mode cooperon and diffuson exist in general
situations (including the Born and unitary limits), the additional
$\pi$-mode cooperon and diffuson appear only in the unitary limit
due to the existence of the particle-hole symmetry, similarly with
the situation of disordered $d$-wave superconductors~\cite{14,15}.
The diffusive 0-modes are gapless, and of diffusive poles at small
energy difference. On the contrary, the diffusive $\pi$-modes are
gapped by the deviation from perfectly-nested Fermi surface
measured by $|\mu|$, with the diffusive poles at the total energy
of $-2\mu$.

In non-unitary cases, only the 0-mode cooperon contributes to the
QI effect, yielding a weak-localization correction to the
conductivity described by Eq.~(1.1). Upon approaching the unitary
limit, there exist some additional conductivity diagrams
containing the $\pi$-mode cooperon or $\pi$-mode diffuson, giving
rise to additional corrections to the conductivity. In the case of
small $|\mu|$, the conductivity in the unitary limit is subject to
power-law anti-localization corrections, which come predominantly
from the contribution of the $\pi$-mode diffuson. however, this
anti-localization effect is suppressed by large deviation from the
nesting case, and the usual logarithmic weak-localization effect
survives due to the 0-mode cooperon. As a result, the dc
conductivity in the unitary limit becomes a non-monotonic function
of the temperature or the sample size.

The calculations of the conductivity at finite temperatures reveal
that, both the inelastic scattering and the thermal fluctuation
are the dephasing factors in the QI processes relevant to the
diffusive $\pi$-modes. In the region of $\gamma_i\gg T,|\mu|$, the
inelastic scattering is the dominant dephasing mechanism. For the
case of $T\gg\gamma_i,|\mu|$, the QI effect is mainly dephased by
the thermal fluctuation. This unique feature is related with the
diffusive poles  of the $\pi$-mode cooperon and diffuson.

At zero temperature, there exists a characteristic length $\xi$
defined by Eq.~(4.42). While the dc conductivity is subject to a
logarithmic weak-localization correction for large samples
($L\gg\xi$), it increases as $L^2$ in the region of $L\ll\xi$.
This non-monotonic behavior of the conductivity might be
instructive for the numerical study of the present system, as the
latter is usually carried out for a finite sample. For a
perfectly-nested  Fermi surface ($\xi\rightarrow\infty$), the
anti-localization effect suggests the existence of extended states
at the band center. We note that the non-localized states were
also found in the center of the band of 2D Anderson model with
purely off-diagonal disorder~[6--10]. Numerical studies of these
systems revealed that the localization length diverges at the band
center~\cite{8,9}. The present analytical work tries to put some
physical insights into the understanding of the delocalization
effect in the considered system.

Since the diffusive $\pi$-modes introduce some additional
conductivity diagrams, they are expected to play non-trivial roles
in other transport properties such as the magnetoresistence and
the Hall effect in this system. How these new diagrams affect the
quasiparticle localization in disordered $d$-wave superconductors
is also an interesting and open problem.

\begin{center}
{\large {\bf Acknowledgment}}
\end{center}

This work was supported by the National Natural Science Foundation
of China under Grants No. 10274008 and No. 10174011, and the
Jiangsu-Province Natural Science Foundation of China under Grant
No. BK2002050. DYX would like to acknowledge the support of Grant
No. G19980614 for State Key Programs for Basic Research of China.
\renewcommand{\theequation}{A\arabic{equation}}
\vspace{0.5cm}
\begin{center}
{\large {\bf Appendix A: Some useful mathematical formulas}}
\end{center}
\setcounter{equation}{0}

In this appendix, we will prove the following mathematical
formulas
\begin{equation}
\sum_{\bm k}\frac{1}{\epsilon_{\bm k}^2+\gamma^2}=\frac{1}{2\pi
a^2t\gamma}\ln\frac{16t}{\gamma},
\end{equation}
\begin{equation}
\sum_{\bm k}(G^R_{\bm k})^m(G^A_{\bm
k})^n=2\pi\rho_0\frac{(m+n-2)!}{(m-1)!(n-1)!}\times\frac{i^{n-m}}{(2\gamma)^{m+n-1}},
\end{equation}
\begin{equation}
\sum_{\bm k}v_{\bm k}^2(G^R_{\bm k})^m(G^A_{\bm
k})^n=4\pi\rho_0D\frac{(m+n-2)!}{(m-1)!(n-1)!}\times\frac{i^{n-m}}{(2\gamma)^{m+n-2}},
\end{equation}
where $m,n\geq1$. Although Eqs.~(A2) and (A3) are exactly the same
as those of the free-electron model, we shall show that they are
also valid for the present nearly half-filled band.

The summation over ${\bm k}$ in Eq.~(A1) can be replaced by an
integral, i.e.,
\begin{equation}
\sum_{\bm k}\frac{1}{\epsilon_{\bm
k}^2+\gamma^2}=\int^{4t}_{-4t}d\epsilon\frac{\rho_{\rm
cl}(\epsilon)}{\epsilon^2+\gamma^2}.
\end{equation}
In the case of $|\mu|\ll\gamma\ll t$ under considered, the density
of states $\rho_{\rm cl}(\epsilon)$ in Eq.~(A4) can be replaced by
the expression near the band center given by Eq.~(2.3), yielding
\begin{eqnarray}
\sum_{\bm k}\frac{1}{\epsilon_{\bm k}^2+\gamma^2}&\approx
&\frac{1}{2\pi^2a^2t}\int\limits^{+\infty}_{-\infty}\frac{d\epsilon}
{\epsilon^2+\gamma^2}\ln\frac{16t}{|\epsilon+\mu|}\nonumber\\
&=&\frac{1}{2\pi^2a^2t\gamma}\int\limits^{+\infty}_{-\infty}\frac{dx}{x^2+1}
\left(\ln\frac{16t}{\gamma}-\ln|x+\frac{\mu}{\gamma}|\right).
\end{eqnarray}
Completing the integral over $x$ in Eq.~(A5), and noting that
\begin{equation}
\int\limits^{+\infty}_{-\infty}dx\frac{\ln|x+\mu/\gamma|}{x^2+1}
\approx\int\limits^{+\infty}_{-\infty}dx\frac{\ln|x|}{x^2+1}=0,
\end{equation}
we immediately obtain Eq.~(A1).

Similarly, a use of Eq.~(2.5) leads to
\begin{eqnarray}
\sum_{\bm k}(G^R_{\bm k})^m(G^A_{\bm k})^n&=&\sum_{\bm
k}\frac{1}{(-\epsilon_{\bm k}+i\gamma)^m(-\epsilon_{\bm
k}-i\gamma)^n}\nonumber\\&\approx
&\frac{(-1)^{m+n}}{2\pi^2a^2t\gamma^{m+n-1}}\int\limits^{+\infty}_{-\infty}dx
\frac{\ln(16t/\gamma)-\ln|x|}{(x-i)^m(x+i)^n}.
\end{eqnarray}
In the case of $\ln(16t/\gamma)\gg1$, the second term in the right
side of Eq.~(A7) can be neglected. Thus, a completion of the
integral over $x$ in Eq.~(A7) leads to Eq.~(A2).

By means of Eqs.~(2.2) and (2.5), we get
\begin{equation}
\sum_{\bm k}v^2_{\bm k}(G^R_{\bm k})^m(G^A_{\bm
k})^n=\int^{4t}_{-4t}\frac{d\epsilon}{(2\pi)^2}\frac{1}{(-\epsilon+i\gamma)^m
(-\epsilon+i\gamma)^n}\int_{\epsilon_{\bm
k}=\epsilon}\frac{ds}{v_{\bm k}}v^2_{\bm k},
\end{equation}
where the integration $\int ds$ is over the constant-energy
surface $\epsilon_{\bm k}=\epsilon$. Since the main contribution
of the right side of Eq.~(A8) comes from the nearly-nested
surfaces of constant energy, we have
\begin{eqnarray}
\sum_{\bm k}v^2_{\bm k}(G^R_{\bm k})^m(G^A_{\bm k})^n&\approx
&\frac{1}{2\pi}\frac{(m+n-2)!}{(m-1)!(n-1)!}\times\frac{i^{n-m}}{(2\gamma)^{m+n-1}}\times
4\int^{\pi/a}_{0}\sqrt{2}dk_x2\sqrt{2}ta\sin k_xa\nonumber\\
&=&\frac{16t}{\pi}\frac{(m+n-2)!}{(m-1)!(n-1)!}\times\frac{i^{n-m}}{(2\gamma)^{m+n-1}}.
\end{eqnarray}
A substitution of Eq.~(3.10) into Eq.~(A9) immediately yields
Eq.~(A3).

\renewcommand{\theequation}{B\arabic{equation}}
\vspace{0.5cm}
\begin{center}
{\large {\bf Appendix B: Vanishing QI correction to dc
conductivity from the RR-correlation function}}
\end{center}
\setcounter{equation}{0}

In this appendix, we will show that the RR-correlation function
from the diagrams with $\pi$-mode cooperon or diffuson contributes
no correction to the dc conductivity.

The contributions of the $\pi$-mode cooperon to the RR-correlation
function come from Figs.~2(a), 2(c), and 2(d), which can be
evaluated as
\begin{eqnarray}
\Pi(\epsilon,\epsilon')^{RR}_{2a}&=&e^2\sum_{{\bm k}{\bm q}}{\bm
v}_{\bm k}\cdot{\bm v}_{\bm k}(G^R_{\bm k})^3G^R_{{\bm Q}-{\bm
k}}C_\pi({\bm
q};\epsilon,\epsilon)^{RR}\nonumber\\&=&-e^2\sum_{{\bm k}{\bm
q}}v^2_{\bm k}(G^R_{\bm k})^3G^A_{\bm k}C_\pi({\bm
q};\epsilon,\epsilon)^{RR},
\end{eqnarray}
\begin{eqnarray}
\Pi(\epsilon,\epsilon')^{RR}_{2c}&=&e^2\sum_{{\bm k}{\bm q}}{\bm
v}_{\bm k}\cdot{\bm v}_{{\bm Q}-{\bm k}}(G^R_{\bm k}G^R_{{\bm
Q}-{\bm k}})^2C_\pi({\bm
q};\epsilon,\epsilon')^{RR}\nonumber\\&=&e^2\sum_{{\bm k}{\bm
q}}v^2_{\bm k}(G^R_{\bm k}G^A_{\bm k})^2C_\pi({\bm
q};\epsilon,\epsilon')^{RR},
\end{eqnarray}
and
\begin{eqnarray}
\Pi(\epsilon,\epsilon')^{RR}_{2d}&=&e^2\sum_{{\bm k}{\bm k'}{\bm
q}}{\bm v}_{\bm k}\cdot{\bm v}_{\bm k'}(G^R_{\bm k}G^R_{\bm
k'})^2G^R_{{\bm Q}+{\bm q}-{\bm k}}G^R_{{\bm Q}+{\bm q}-{\bm
k'}}C_\pi({\bm q};\epsilon,\epsilon)^{RR}C_\pi({\bm
q};\epsilon,\epsilon')^{RR}\nonumber\\&=&\frac{e^2}{4}\sum_{{\bm
k}{\bm k'}{\bm q}}q^2v^2_{\bm k}v^2_{\bm k'}(G^R_{\bm k}G^A_{\bm
k}G^R_{\bm k'}G^A_{\bm k'})^2C_\pi({\bm
q};\epsilon,\epsilon)^{RR}C_\pi({\bm q};\epsilon,\epsilon')^{RR}.
\end{eqnarray}
A use of Eqs.~(A2) and (A3) into Eqs.~(B1)--(B3) immediately
yields
\begin{equation}
\Pi(\epsilon,\epsilon')^{RR}_{2a}=-\sum_{\bm
q}\frac{2e^2D}{Dq^2-i2(\epsilon+\mu)+2\gamma_i},
\end{equation}
\begin{equation}
\Pi(\epsilon,\epsilon')^{RR}_{2c}=-\sum_{\bm
q}\frac{4e^2D}{Dq^2-i(\epsilon+\epsilon'+2\mu)+2\gamma_i},
\end{equation}
and
\begin{equation}
\Pi(\epsilon,\epsilon')^{RR}_{2d}=\sum_{\bm
q}\frac{4e^2D^2q^2}{[Dq^2-i2(\epsilon+\mu)+2\gamma_i]
[Dq^2-i(\epsilon+\epsilon'+2\mu)+2\gamma_i]}.
\end{equation}

The contributions of the $\pi$-mode diffuson to the RR-correlation
function come from Figs.~3(a) and 3(c)--3(e), which can be
similarly calculated as
\begin{eqnarray}
\Pi(\epsilon,\epsilon')^{RR}_{3a}&=&e^2\sum_{{\bm k}{\bm q}}{\bm
v}_{\bm k}\cdot{\bm v}_{\bm k}(G^R_{\bm k})^3G^R_{{\bm Q}+{\bm
k}}W^{RR}\Lambda^R({\bm q},\epsilon)^2\nonumber\\&=&-e^2\sum_{{\bm
k}{\bm q}}v^2_{\bm k}(G^R_{\bm k})^3G^A_{\bm
k}W^{RR}\Lambda^R({\bm q},\epsilon)^2,
\end{eqnarray}
\begin{eqnarray}
\Pi(\epsilon,\epsilon')^{RR}_{3c}&=&e^2\sum_{{\bm k}{\bm q}}{\bm
v}_{\bm k}\cdot{\bm v}_{{\bm Q}+{\bm k}}(G^R_{\bm k}G^R_{{\bm
Q}+{\bm k}})^2W^{RR}\Lambda^R({\bm q},\epsilon)\Lambda^R({\bm
q},\epsilon')\nonumber\\&=&-e^2\sum_{{\bm k}{\bm q}}v^2_{\bm
k}(G^R_{\bm k}G^A_{\bm k})^2W^{RR}\Lambda^R({\bm
q},\epsilon)\Lambda^R({\bm q},\epsilon'),
\end{eqnarray}
\begin{eqnarray}
\Pi(\epsilon,\epsilon')^{RR}_{3d}&=&e^2\sum_{{\bm k}{\bm k'}{\bm
q}}{\bm v}_{\bm k}\cdot{\bm v}_{\bm k'}(G^R_{\bm k}G^R_{\bm
k'})^2G^R_{{\bm Q}+{\bm q}+{\bm k}}G^R_{{\bm Q}+{\bm q}+{\bm
k'}}W^{RR}\Lambda^R({\bm q},\epsilon)^2D_\pi({\bm
q};\epsilon,\epsilon')^{RR}\nonumber\\&=&\frac{e^2}{4}\sum_{{\bm
k}{\bm k'}{\bm q}}q^2v^2_{\bm k}v^2_{\bm k'}(G^R_{\bm k}G^A_{\bm
k}G^R_{\bm k'}G^A_{\bm k'})^2W^{RR}\Lambda^R({\bm
q},\epsilon)^2D_\pi({\bm q};\epsilon,\epsilon')^{RR},
\end{eqnarray}
and
\begin{eqnarray}
\Pi(\epsilon,\epsilon')^{RR}_{3e}&=&e^2\sum_{{\bm k}{\bm k'}{\bm
q}}{\bm v}_{\bm k}\cdot{\bm v}_{\bm k'}(G^R_{\bm k}G^R_{\bm
k'})^2G^R_{{\bm Q}+{\bm q}+{\bm k}}G^R_{-{\bm Q}-{\bm q}+{\bm
k'}}W^{RR}\Lambda^R({\bm q},\epsilon)\Lambda^R({\bm
q},\epsilon')D_\pi({\bm
q};\epsilon,\epsilon')^{RR}\nonumber\\&=&-\frac{e^2}{4}\sum_{{\bm
k}{\bm k'}{\bm q}}q^2v^2_{\bm k}v^2_{\bm k'}(G^R_{\bm k}G^A_{\bm
k}G^R_{\bm k'}G^A_{\bm k'})^2W^{RR}\Lambda^R({\bm
q},\epsilon)\Lambda^R({\bm q},\epsilon')D_\pi({\bm
q};\epsilon,\epsilon')^{RR},
\end{eqnarray}
leading to
\begin{equation}
\Pi(\epsilon,\epsilon')^{RR}_{3a}=-\sum_{\bm
q}\frac{4e^2D\gamma}{[Dq^2-i2(\epsilon+\mu)+2\gamma_i]^2},
\end{equation}
\begin{equation}
\Pi(\epsilon,\epsilon')^{RR}_{3c}=\sum_{\bm
q}\frac{8e^2D\gamma}{[Dq^2-i2(\epsilon+\mu)+2\gamma_i]
[Dq^2-i2(\epsilon'+\mu)+2\gamma_i]},
\end{equation}
\begin{equation}
\Pi(\epsilon,\epsilon')^{RR}_{3d}=\sum_{\bm
q}\frac{8e^2D^2q^2\gamma}{[Dq^2-i2(\epsilon+\mu)+2\gamma_i]^2
[Dq^2-i(\epsilon+\epsilon'+2\mu)+2\gamma_i]},
\end{equation}
and
\begin{equation}
\Pi(\epsilon,\epsilon')^{RR}_{3e}=-\sum_{\bm
q}\frac{8e^2D^2q^2\gamma}{[Dq^2-i2(\epsilon+\mu)+2\gamma_i][Dq^2-i2(\epsilon'+\mu)+2\gamma_i]
[Dq^2-i(\epsilon+\epsilon'+2\mu)+2\gamma_i]}.
\end{equation}

Substituting Eqs.~(B4)--(B6) into Eqs.~(4.3) and (4.34), one can
easily show that
\begin{equation}
2\sigma(T)^{RR}_{2a}=\sigma(T)^{RR}_{2c}=-\sigma(T)^{RR}_{2d}
=\frac{e^2}{8\pi^2}\int\limits_{-\infty}^{+\infty}\frac{dx}{\cosh^2x}
\ln\frac{\gamma^2}{(2Tx+\mu)^2+\gamma_i^2}
\end{equation}
and
\begin{equation}
2\sigma(L)^{RR}_{2a}=\sigma(L)^{RR}_{2c}=-\sigma(L)^{RR}_{2d}=\frac{e^2}{4\pi^2}
\ln\frac{\gamma^2}{\gamma_L^2+\mu^2},
\end{equation}
leading to
\begin{equation}
\sigma^{RR}_{\pi-{\rm
coop}}=2\sigma^{RR}_{2a}+\sigma^{RR}_{2c}+2\sigma^{RR}_{2d}=0.
\end{equation}
From Eqs.~(4.3) and (B11)--(B14), it follows that
\begin{equation}
2\sigma^{RR}_{3a}=-\sigma^{RR}_{3c}~{\rm
and}~\sigma^{RR}_{3d}=-\sigma^{RR}_{3e},
\end{equation}
yielding
\begin{equation}
\sigma^{RR}_{\pi-{\rm
diff}}=2\sigma^{RR}_{3a}+\sigma^{RR}_{3c}+2\sigma^{RR}_{2d}+2\sigma^{RR}_{2d}=0.
\end{equation}
Equations (B17) and (B19) indicate that the RR-correlation
function from the contribution of the $\pi$-mode cooperon or
diffuson has a vanishing correction to the dc conductivity.

\newpage

\end{document}